%% file: Payne_Mallick_TwoStageMH.tex
\documentclass[12pt]{article}
\usepackage{geometry}
\usepackage{amsmath}
\usepackage{setspace}
\usepackage{graphicx}
\usepackage{epstopdf}
\usepackage{natbib}

\newcommand{\bbeta}[0]{\boldsymbol{\beta}}
\newcommand{\Bbeta}[0]{\boldsymbol{\beta}}
\newcommand{\Btheta}[0]{\boldsymbol{\theta}}
\newcommand{\Bx}[0]{\mathbf{X}}
\newcommand{\bx}[0]{\mathbf{x}}
\newcommand{\by}[0]{\mathbf{y}}

\begin{document}
\bibpunct{(}{)}{,}{a}{}{;}
\begin{spacing}{2}
%\runninghead{R.\ D.\ Payne and B.\ K.\ Mallick}{Two-stage metropolis-hastings for tall data}
\title{Two-Stage Metropolis-Hastings for Tall Data}
\author{Richard D.\ Payne, Bani K.\ Mallick}
%\address{\affilnum{a}Department of Statistics, Texas A\&M University, College Station, TX 77843, USA}
%\corremail{richard@stat.tamu.edu}
%\received{00 Month Year}
%\accepted{00 Month Year}
\maketitle
\begin{abstract}
This paper discusses the challenges presented by tall data problems associated with Bayesian classification (specifically binary classification) and the existing methods to handle them.  Current methods include parallelizing the likelihood, subsampling, and consensus Monte Carlo. A new method based on the two-stage Metropolis-Hastings algorithm is also proposed. The purpose of this algorithm is to reduce the exact likelihood computational cost in the tall data situation. In the first stage, a new proposal is tested by the approximate likelihood based model. The full likelihood based posterior computation will be conducted only if the proposal passes the first stage screening. Furthermore, this method can be adopted into the consensus Monte Carlo framework.  The two-stage method is applied to logistic regression, hierarchical logistic regression, and Bayesian multivariate adaptive regression splines.

\end{abstract}

\section{Introduction}
In the past twenty-five years, Bayesian statistics have become increasingly popular as they are capable of analyzing data with complex structures. Consequently, Bayesian methods have been proven to be effective in a wide range of applications.  The rise in popularity is largely attributed to simulation based algorithms which can approximate the complex posterior distributions of non-conjugate models, such as Markov Chain Monte Carlo (MCMC) methods including the Metropolis-Hastings (MH)  algorithm~\citep*{robert2013monte}.

The term ``tall data" generally describes data in which $n >> p$, that is, when the number of observations is much larger than the number of predictors.  For MCMC methods, as $n$ increases, so does the computational demand of the algorithm.  Specifically, for MH, the increased computational demand is driven by the complete scan of the data through likelihood evaluations on each iteration of the algorithm.  If $n$ is large enough, MCMC  methods (including MH) are computationally infeasible.

There are several general methods to overcome this issue.  The simplest method involves parallelizing the likelihood to speed up computation.  Another method divides the data across multiple machines and performs independent parallel MCMC on each machine to sample from the posterior distribution (consensus Monte Carlo).  The results are then aggregated using weighting~\citep*{scott}.  A third approach is to use subsampling methods to provide a faster estimation of the likelihood~\citep*{quiroz, korattikara, bardenet}.  See~\citet*{bardenet2015markov} for a review of MCMC approaches for tall data.

We propose a method based on a two-stage Metropolis algorithm which uses a cheap estimate of the likelihood to determine if a full estimation of the likelihood is necessary. Furthermore, we compare the strengths and weaknesses of the general tall data MH methods of consensus, subsampling, and two-stage Metropolis, as well as briefly introduce the use of a combination of the consensus and two-stage methods.  For definiteness, in the following, the focus of this paper is on the classification problem. However, the developed methodology can be extended to any model which is suitable to analyze tall data.  The methods are applied to three datasets: marketing data from a Portuguese bank, loan data from Freddie Mac, and a simulated dataset.  Logistic regression is applied to both the Portuguese bank and Freddie Mac datasets and an additional logistic hierarchical model is fit to the Freddie Mac dataset.  Modifications to the techniques described in the papers above have been made to accommodate the features of these datasets and are explained further.  Lastly, the two-stage method is applied to a simulated binary classification problem using Bayesian multivariate adaptive regression splines (BMARS).

In Section 2, we describe the existing methods for handling tall data and present the two-stage methodology.  Section 3 applies the methods to the marketing, Freddie Mac, and simulated datasets.  Section 4 provides a brief discussion and Section 5 concludes.

\section{Methodology}

We begin by briefly describing existing techniques to speed up MCMC computation for tall data applications.

\subsection{Likelihood Parallelization}
Perhaps the simplest way to adapt the Metropolis-Hastings algorithm for tall data is to compute the likelihood in parallel.  In this method, the data are partitioned into $p$ partitions and each is assigned to a separate process/core.  On each iteration of the MH algorithm, the master process draws parameters from the proposal distribution and sends the proposed values to the other processes.  Each process then computes the likelihood for its partition and passes this information (i.e.\ the sum of the log-likelihood) to the master process which sums the log-likelihood contributions from each partition and determines whether or not to accept the proposal.  As long as there is no significant communication overhead between the processes, the MH algorithm's speed will be increased while still sampling from the true posterior distribution.

\subsection{Consensus Monte Carlo}
In the consensus Monte Carlo method the data are randomly partitioned into $p$ partitions. Subsequently, allow each partition to run a full MCMC simulation from a posterior distribution given its own data. Lastly, combine the posterior simulations from each partition to produce a set of global draws to reproduce the unified posterior distribution.

Suppose $\by=(y_{1},\ldots ,y_{n})$ denotes the full data and $\by_{j}$  is the data at the $j$th partition. We then represent the posterior distribution of $\bbeta$ as 
\begin{displaymath}
p(\bbeta|\by)\propto \prod_{j=1}^{p}p(\by_{j}|\bbeta){p(\bbeta)}^{1/p}
\end{displaymath}

where the prior distribution has been expressed as the product of the $p$ components. 

    For each partition, a Metropolis sampler with a chain of length $m$ is computed in parallel with the prior weight adjusted to $p^{-1}$ its original weight.  Once posterior samples are obtained from each of the partitions, the results are combined using a weighted average.  The weight, $W_i$,  for the $i$th partition is equal to the inverse of the posterior covariance matrix obtained from the Metropolis sampler.  Let $\bbeta_i$ be the posterior sample matrix from the $i$th partition.  Thus, the final posterior sample, $\bbeta$ is obtained using the following weighted average: 

\begin{eqnarray*}
\bbeta =\left( \sum_{i=1}^p \bbeta_iW_i \right)  \left(\sum_{j=1}^p W_j\right)^{-1} 
\end{eqnarray*}
For details see~\citet{scott}.

\subsection{Subsampling Based Methods}
In subsampling methods, a small subset of the data is used to estimate the likelihood function which is then used to evaluate the acceptance probabilities of the MH algorithm.  In principle, subsampling reduces the data size and therefore a faster MCMC algorithm can be developed.  Using an unbiased likelihood estimate in the MCMC chain still provides the correct stationary distribution~\citep*{andrieu2009}, however the efficiency of the MCMC chain depends on the variance of the estimator.  Usually complete random sampling does not work well in this situation (i.e.\ the chain gets stuck for many iterations), but some general guidelines for estimating the full likelihood from a subsample in a Bayesian setting have been developed.~\citet{quiroz} suggest using a portion of the data prior to MCMC and fitting Gaussian processes or splines to approximate the log-likelihood.  On each iteration of the MCMC chain, the log-likelihood is estimated for each observation.  The data are then sampled with probability proportional to its estimated log-likelihood value, which reduces the variance of the estimator and improves the efficiency of the chain. 

\subsection{Two-Stage MH}
Consider the usual Bayesian model setup where the posterior distribution of the parameter $\bbeta$ given data $\by$ is given by
\begin{equation}
p(\bbeta|\by)\propto p(\by|\bbeta)p(\bbeta)
\end{equation}
where  $p(\by|\bbeta)$ is the likelihood function and $p(\bbeta)$ is the prior distribution for the parameter vector $\bbeta$. If a non-conjugate prior is selected, the posterior distribution $p(\bbeta|\by)$ often cannot be expressed in an explicit form and consequently MCMC methods must be used to simulate samples from this posterior distribution. More specifically, we use the Metropolis-Hastings (MH) algorithm to generate samples of $\bbeta$s from $p(\bbeta|\by)$. The MH algorithm is described as follows.

{\bf A.1 MH Algorithm}
\begin{enumerate}
\item At the $t$th iteration generate $\bbeta$ from the proposal distribution $q(\bbeta|\bbeta_{t})$ where $\bbeta_{t}$ is the current state
\item Accept $\bbeta$ as a posterior sample with probability
\begin{equation}
h(\bbeta_{t},\bbeta)=\text{min}\left\{1,\frac{q(\bbeta_{t}|\bbeta)p(\bbeta|\by)}{q(\bbeta|\bbeta_{t})p(\bbeta_{t}|\by)}\right\}
\end{equation}
\item $\bbeta_{t+1}=\bbeta$ with probability $h(\bbeta_{t},\bbeta)$ and $\bbeta_{t+1}=\bbeta_{t}$ with probability $1-h(\bbeta_{t},\bbeta)$.
\end{enumerate}
At each iteration, the probability of moving from the state $\bbeta_{t}$ to next state $\bbeta$ is $q(\bbeta|\bbeta_{t})h(\bbeta_{t},\bbeta)$, hence the transition kernel for the Markov Chain $\bbeta_{t}$ is
\begin{displaymath}
T(\bbeta_{t},\bbeta)=q(\bbeta|\bbeta_{t})h(\bbeta_{t},\bbeta)+\left\{1-\int{q(\bbeta|\bbeta_{t})h(\bbeta_{t},\bbeta)d\bbeta}\right\}I(\bbeta=\bbeta_{t})
\end{displaymath} 
where $I()$ is the indicator function.
Due to the iterative nature of the algorithm, the likelihood function $p(\by|\bbeta)$ needs to be evaluated repeatedly which is expensive when $n$ is large. Hence, we need to modify the MH algorithm to adapt it for tall data problems.

In the MH algorithm described in A.1, the  evaluation of the likelihood is expensive in the tall data situation. Generally the MCMC chain requires thousands of iterations to converge. Furthermore, we need to generate a large number of samples to quantify the uncertainty in the parameters.  We use the two-stage MH algorithm where the proposal distribution $q()$ is adapted to the target distribution using an approximate likelihood based model. These algorithms  have been used previously~\citep*{christen,higdon,mondal}, usually for solving expensive inverse problems.  For our purposes, instead of testing each proposal by the exact likelihood based model directly, initially  the algorithm tests the proposal by the approximate likelihood based model which is much cheaper to compute. If the proposal is accepted by the initial test, then an exact likelihood based computation will be conducted and the proposal will be further tested as in the MH algorithm method described in A.1. Otherwise, the proposal will be rejected by the approximate model and a new proposal will be generated from $q()$. The approximate likelihood based model filters the unacceptable proposals and avoids the expensive full likelihood computations.

{\bf A.2 Two-Stage MH Algorithm}

Let $\hat{p}(\by|\bbeta)$ be an approximation of the full likelihood, and let the approximate posterior distribution be represented as $p^{*}(\bbeta|\by)\propto \hat{p}(\by|\bbeta)p(\bbeta)$.  Then the Two-Stage MH Algorithm proceeds as follows: 

\begin{enumerate}
\item At the $t$th iteration generate $\bbeta^{'}$ from the proposal distribution $q(\bbeta^{'}|\bbeta_{t})$
\item Take a real proposal as
\[
 \bbeta =
  \begin{cases}
   \bbeta^{'} & \text{with probability } \delta(\bbeta_{t},\bbeta^{'})\\
   \bbeta_{t}     & \text{with probability } 1-\delta(\bbeta_{t},\bbeta^{'})
  \end{cases}
\]
where 
\begin{displaymath}
\delta(\bbeta_{t},\bbeta^{'})=\text{min}\left\{1,\frac{q(\bbeta_{t}|\bbeta^{'})p^{*}(\bbeta^{'}|\by)}{q(\bbeta^{'}|\bbeta_{t})p^{*}(\bbeta_{t}|\by)}\right\}
\end{displaymath}

\item Accept $\bbeta$ as a posterior sample with probability
\begin{equation}
\rho(\bbeta_{t},\bbeta)=\text{min}\left\{1,\frac{Q(\bbeta_{t}|\bbeta)p(\bbeta|\by)}{Q(\bbeta|\bbeta_{t})p(\bbeta_{t}|\by)}\right\}
\end{equation}
where $Q(\bbeta|\bbeta_{t})=\delta(\bbeta_{t},\bbeta)q(\bbeta|\bbeta_{t})+\{1-\int{\delta(\bbeta_{t},\bbeta)q(\bbeta|\bbeta_{t})d\bbeta}\}I(\bbeta=\bbeta_{t})$
\item Hence take $\bbeta_{t+1}=\bbeta$ with probability $\rho(\bbeta_{t},\bbeta)$ and $\bbeta_{t+1}=\bbeta_{t}$ with probability $1-\rho(\bbeta_{t},\bbeta)$.
\end{enumerate}
At each iteration, the probability of moving from the state $\bbeta_{t}$ to next state $\bbeta$ is $q(\bbeta|\bbeta_{t})\rho(\bbeta_{t},\bbeta)$, hence the transition kernel for the Markov Chain $\bbeta_{t}$ is
\begin{displaymath}
T(\bbeta_{t},\bbeta)=q(\bbeta|\bbeta_{t})\rho(\bbeta_{t},\bbeta)+\left\{1-\int{q(\bbeta|\bbeta_{t})\rho(\bbeta_{t},\bbeta)d\bbeta}\right\}I(\bbeta=\bbeta_{t}).
\end{displaymath} 
In the above algorithm, if the trial proposal $\bbeta^{'}$ is rejected by the approximate posterior then no further computation is needed. Thus, the expensive exact posterior computation can be avoided for those proposals which are unlikely to be accepted. This is just an adaption of the proposal using the approximate posterior where the transition kernel can be written as $K(\bbeta_{t},\bbeta)=\rho(\bbeta_{t},\bbeta)Q(\bbeta|\bbeta_{t})$ for $\bbeta\neq \bbeta_{t}$ and $K(\bbeta_{t},\{\bbeta_{t}\})=1-\int_{\bbeta\neq\beta_{t}}{\rho(\bbeta_{t},\bbeta)Q(\bbeta_{t}|\bbeta)d\bbeta}$ for $\bbeta=\bbeta_{t}$. It is simple to show that the detailed balance condition $p(\bbeta_{t}|\by)K(\bbeta_{t},\bbeta)=p(\bbeta|\by)K(\bbeta,\bbeta_{t})$ is always satisfied under some minor regularity conditions like the regular MH algorithm. 

{\bf Result 1}: The detailed balance condition is satisfied under the regularity conditions of the MH algorithm.  That is, $p(\bbeta_{t}|\by)K(\bbeta_{t},\bbeta)=p(\bbeta|\by)K(\bbeta,\bbeta_{t})$.

{\bf Proof}: When $\bbeta = \bbeta_t$, the result is trivial.  When $\bbeta \neq \bbeta_t$ we have 
\begin{eqnarray*}
p(\bbeta_{t}|\by)K(\bbeta_{t},\bbeta)&=&p(\bbeta_{t}|\by)\rho(\bbeta_t,\bbeta)Q(\bbeta|\bbeta_t) \\
&=& p(\bbeta_{t}|\by)\min \left\{1, \frac{Q(\bbeta_t | \bbeta)p(\bbeta |\by)}{Q(\bbeta |\bbeta_t )p(\bbeta_t |\by)} \right\} Q(\bbeta |\bbeta_t)\\
&=& \min\left\{ p(\bbeta_{t}|\by)Q(\bbeta |\bbeta_t), Q(\bbeta_t | \bbeta) p(\bbeta |\by)\right\} \\
&=&  \min\left\{ \frac{p(\bbeta_{t}|\by)Q(\bbeta |\bbeta_t)}{p(\bbeta |\by)Q(\bbeta_t | \bbeta)}, 1 \right\} p(\bbeta |\by)Q(\bbeta_t | \bbeta)\\
&=& \rho(\bbeta,\bbeta_t)p(\bbeta |\by)Q(\bbeta_t | \bbeta) \\
&=& p(\bbeta|\by)K(\bbeta,\bbeta_t)
\end{eqnarray*}

{\bf Result 2}: The acceptance probability can be expressed as 
$$\rho(\bbeta_{t},\bbeta)=\text{min}\left\{1,\frac{p^{*}(\bbeta_{t}|\by)p(\bbeta|\by)}{p^{*}(\bbeta|\by)p(\bbeta_{t}|\by)}\right\}$$

{\bf Proof}: If $\bbeta=\bbeta_{t}$ then the result is trivial since $\rho(\bbeta_{t},\bbeta)=1$. For $\bbeta\neq \bbeta_{t}$
\begin{eqnarray}
Q(\bbeta_{t}|\bbeta)&=&\delta(\bbeta,\bbeta_{t})q(\bbeta_{t}|\bbeta)\nonumber\\
&=&\frac{1}{p^{*}(\bbeta|\by)}\text{min}\{q(\bbeta_{t}|\bbeta)p^{*}(\bbeta|\by),q(\bbeta|\bbeta_{t})p^{*}(\bbeta_{t}|\by)\}\nonumber\\
&=&\frac{q(\bbeta|\bbeta_{t})p^{*}(\bbeta_{t}|\by)}{p^{*}(\bbeta|\by)}\delta(\bbeta_{t},\bbeta)\nonumber \\
&=&\frac{p^{*}(\bbeta_{t}|\by)}{p^{*}(\bbeta|\by)}Q(\bbeta|\beta_{t}).\nonumber
\end{eqnarray}
Substituting this in the expression of $\rho(\bbeta_{t},\bbeta)$ we obtain the required expression.

It is important to note that the methodology above is general enough to be applied to any computationally expensive MH sampler.  However, for definiteness in following, the method is applied to a few specific classification models.  The success of the two-stage method in any given model depends on the construction of a computationally cheap and accurate estimate of the likelihood.  The accuracy and speed of the likelihood estimator governs the efficiency of the MCMC chain.  For instance, if the likelihood estimator $\hat{p}(\by|\bbeta^\prime)$ severely underestimates $p(\by|\bbeta^\prime)$, then $\delta(\bbeta_t,\bbeta^\prime)$ will be small and the proposal will be rejected (even if it might be a reasonable candidate).  On the other hand, if $\hat{p}(\by|\bbeta^\prime)$ severely overestimates $p(\by|\bbeta^\prime)$, then it will likely pass the first stage and get rejected in the second stage since $\rho(\bbeta_t,\bbeta)$ decreases as a function of $p^*(\bbeta|\by) = \hat{p}(\by|\bbeta)p(\bbeta)$; thus the algorithm will compute the full likelihood for an unfavorable candidate.  Consequently, it is important to select an accurate approximation to the likelihood.  Specific likelihood approximations will be discussed for the examples in Section 3.

\subsection{Combining Consensus with Two-Stage MH}
For larger data sets which may not fit in RAM, we propose a combination of the consensus and the two-stage Metropolis methods.  This is identical to the consensus method with the exception that each partition uses the two-stage Metropolis sampler rather than the usual Metropolis sampler.  Since two-stage MH will draw from the same distribution as MH on each partition, the results of the consensus method will remain the same.

\section{Applications}

The methods introduced above were implemented on three datasets.  For initial testing, the methods were implemented on a relatively small dataset with just over 40,000 observations from a phone marketing campaign conducted by a Portuguese bank.  A larger dataset of approximately 2.3 million observations consisting of individual household loan data from Freddie Mac was used to test how the methods scale.  In both, logistic regression was used to classify observations, the latter also employs a hierarchical model.  Lastly, the two-stage method was implemented on a BMARS model with a large ($10^6$ observations) simulated dataset.

\subsection{Logistic Regression Model}

We are considering a binary classification problem where the response $\by$ takes the value 0 or 1 where $\by=(y_{1},\ldots, y_{n})$ and we have a vector of covariates $\bx$. We use a logit link function to link the $i$th response with the covariates as
\begin{eqnarray*}
y_i \ |\ \bbeta, \bx_i &\sim& \text{Bernoulli}\{\pi(\bx_i)\}\\
\pi(\bx_i) &=& \{1 + \text{exp}({-\bx_i\boldsymbol{\beta}})\}^{-1}\\
\boldsymbol{\beta} &\sim& \text{Multivariate-Normal}(\mathbf{0},\Sigma_0) 
\end{eqnarray*}
where $\bbeta$ is the $l$ dimensional vector of classification parameters, $\bx_i$ is the $i$th row of the design matrix ($i=1,\ldots,n$), and  a Gaussian prior is placed on $\bbeta$. The model's posterior distribution can be expressed as 
$p(\bbeta|\by, \bx)\propto p(\by|\bx, \bbeta)p(\bbeta)$.

In the logistic regression models for both the Portuguese bank and Freddie Mac datasets, we estimate the log-likelihood using a variant of the case-control approximate likelihood~\citep*{raftery}. To understand the approximation, it is important to realize the log-likelihood for a logistic regression model can be written as two sums:
\begin{equation}\label{eq:loglike}
\text{log}\{p(\by|\bbeta,\bx)\} = \sum_{i:y_i=1}    \big\{\theta_{i}-\text{log}(1+e^{\theta_{i}})\big\}+ \sum_{i:y_i=0}{-\text{log}(1+e^{\theta_{i}})}
\end{equation}
where $\theta_{i}=\text{log}\{\pi(\bx_{i})/[1-\pi(\bx_{i})]\}=\bx_{i}\bbeta$.

If the data are sparse, then the computation of the first sum will be relatively cheap, and only the second summation needs to be estimated. We use a  subsampling method where a random sample of $a$ observations is taken from the failed outcomes (i.e. $y_i = 0$).  The second sum in (\ref{eq:loglike}) is estimated by multiplying the average log-likelihood of the $a$ sampled observations by $n_0 = \sum_{i=1}^n I(y_i = 0)$, the number of failures in the dataset. Let $A$ be the index values of the subsample of size $a$.  Thus, the original log-likelihood is estimated as

\begin{equation}
\widehat{\text{log}}\{p(\by|\bbeta,\bx)\}= \sum_{i:y_i=1}  \big\{\theta_{i}-\text{log}(1+e^{\theta_{i}})\big\}+ \frac{n_0}{a}\sum_{i\in A}{-\text{log}(1+e^{\theta_{i}})}.
\end{equation}

We note $\widehat{\text{log}}\{p(\by|\bbeta,\bx)\}$ is an unbiased estimate of the log-likelihood as $E[\widehat{\text{log}}\{p(\by|\bbeta,\bx)\}]=\text{log}\{p(\by|\bbeta,\Bx)\}$.  We could obtain an unbiased estimate of the likelihood by making a bias correction~\citep{quiroz} but that is not necessary for our method as we are doing further filtering of the proposal  using the exact likelihood method.

The above approximation to the likelihood yields the following result:

{\bf Result 3}: When the proposal $\beta$ is promoted from the first stage, then for large  $a$ it can be shown that $\rho(\cdot,\cdot)$ goes towards 1. Thus, the two-stage MH algorithm only calculates the original full data likelihood when there is a high probability of acceptance of the proposal.

\subsubsection{Portuguese Bank Data}

The Portuguese bank dataset was obtained from the University of California Irvine Machine Learning Archive and were analyzed in a recent paper by~\citet*{moro}. The binary dependent variable of interest was whether or not a client subscribed to a term deposit after contact through a telephone marketing campaign.  The predictor variables of interest were the client's previous promotion outcome (non-existent, failure, success), age (years), and type of contact (telephone, cellular), education level (8 categories), and marital status (married, divorced, single, unknown).  The categorical variables were treated as nominal values and the continuous variable age was logged and centered.  A vague prior was placed on $\bbeta$ resulting in the following model:
\begin{eqnarray*}
y_i\ |\ \bbeta, \bx_i &\sim& \text{Bernoulli}\{\pi(\bx_i)\}\\
\pi(\bx_i) &=& \{1 + \text{exp}(-\bx_i\boldsymbol{\beta})\}^{-1}\\
\boldsymbol{\beta} &\sim& \text{Multivariate-Normal}(\mathbf{0},\Sigma_0 = I * 10^2)
\end{eqnarray*}
where $\Bbeta$ is the vector of coefficients, $\Sigma_0$ is the prior covariance matrix for $\Bbeta$, $I$ is the identity matrix and $\bx_i$ is the $i$th row of the design matrix, $i=1,\ldots,n$.

The two-stage, consensus, and standard MH algorithms were coded in Fortran and run for 100,000 iterations with a burn in of 5,000 values.  The subsampling method was considerably slower and was consequently run for only 10,000 iterations with the same burn in of 5,000.  In the consensus method, the data were randomly split into 14 partitions.  In the two-stage method, a single random subsample of 1,400 observations was taken prior to MCMC.  This subsample was used to approximate the log-likelihood during the first stage on each iteration of the two-stage MH algorithm.

In the subsampling method, a thin-plate spline surface was fit to a subsample of the data (1,000 observations) prior to MCMC.  For simplicity, in this smaller dataset, the thin-plate spline surface treated the categorical predictors as continuous.  Although this resulted in a somewhat crude approximation to the log-likelihood surface, using a subsample of around 7,000 observations on each iteration allowed the MCMC chain to mix satisfactorily.  To get a better idea of the speed of the subsampling method if a better spline surface was fit, it was also run subsampling 100 observations rather than 7,000.  The number of likelihood evaluations per second for the MH and subsampling method (100 and 7,000 observations) were 355, 23.9, and 1.8 respectively.

\begin{figure}
\centering
%\captionsetup{format=hang}
\includegraphics[scale=.5,angle=270]{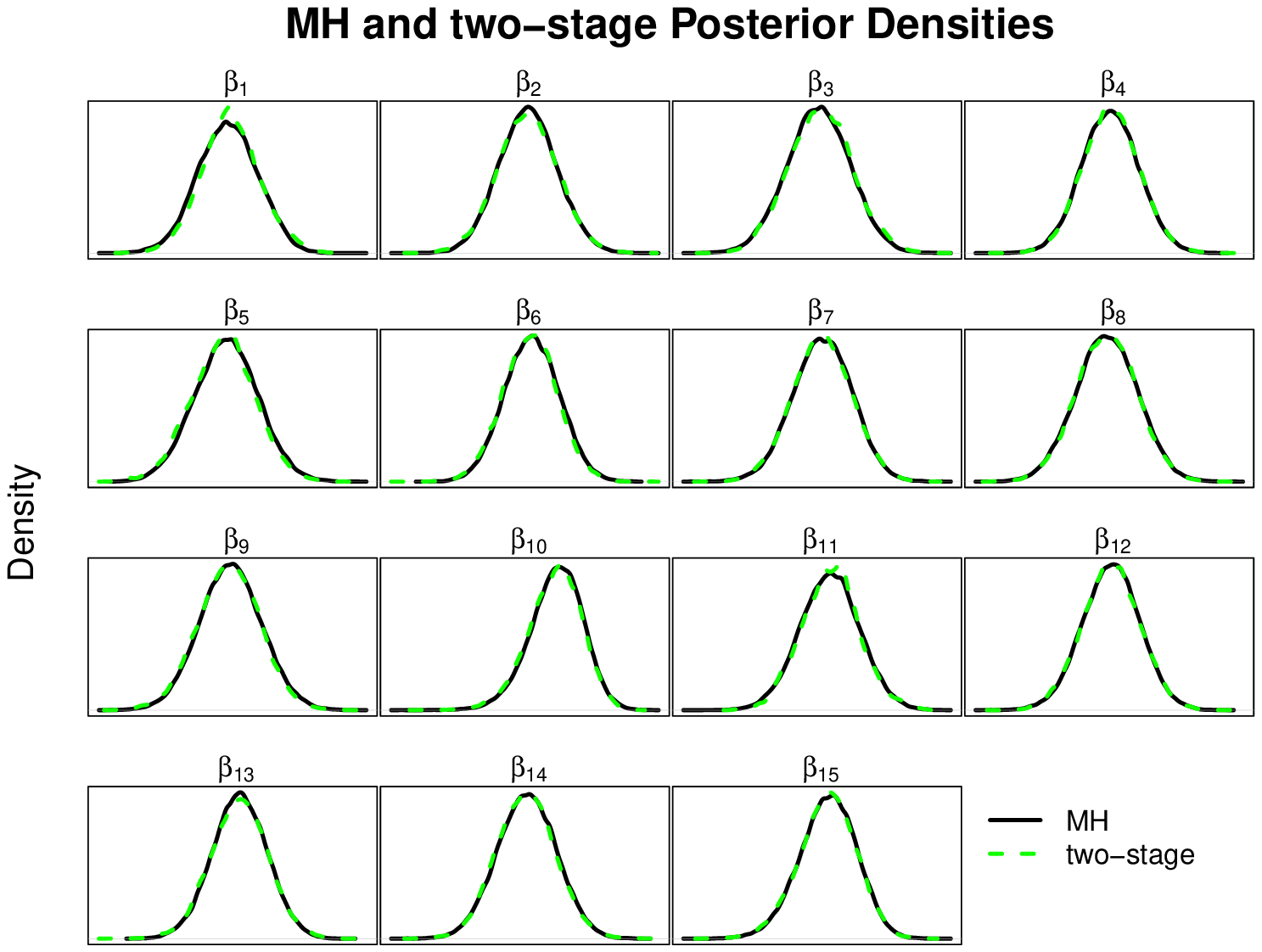}
\caption{Posterior densities from the Portuguese bank data, two-stage Metropolis vs. Metropolis-Hastings.  The MH and two-stage MH methods are represented by the solid and dashed lines, respectively.}
\label{fig:banktwostage}
\end{figure}

\begin{figure}
\centering
%\captionsetup{format=hang}
\includegraphics[scale=.5,angle=270]{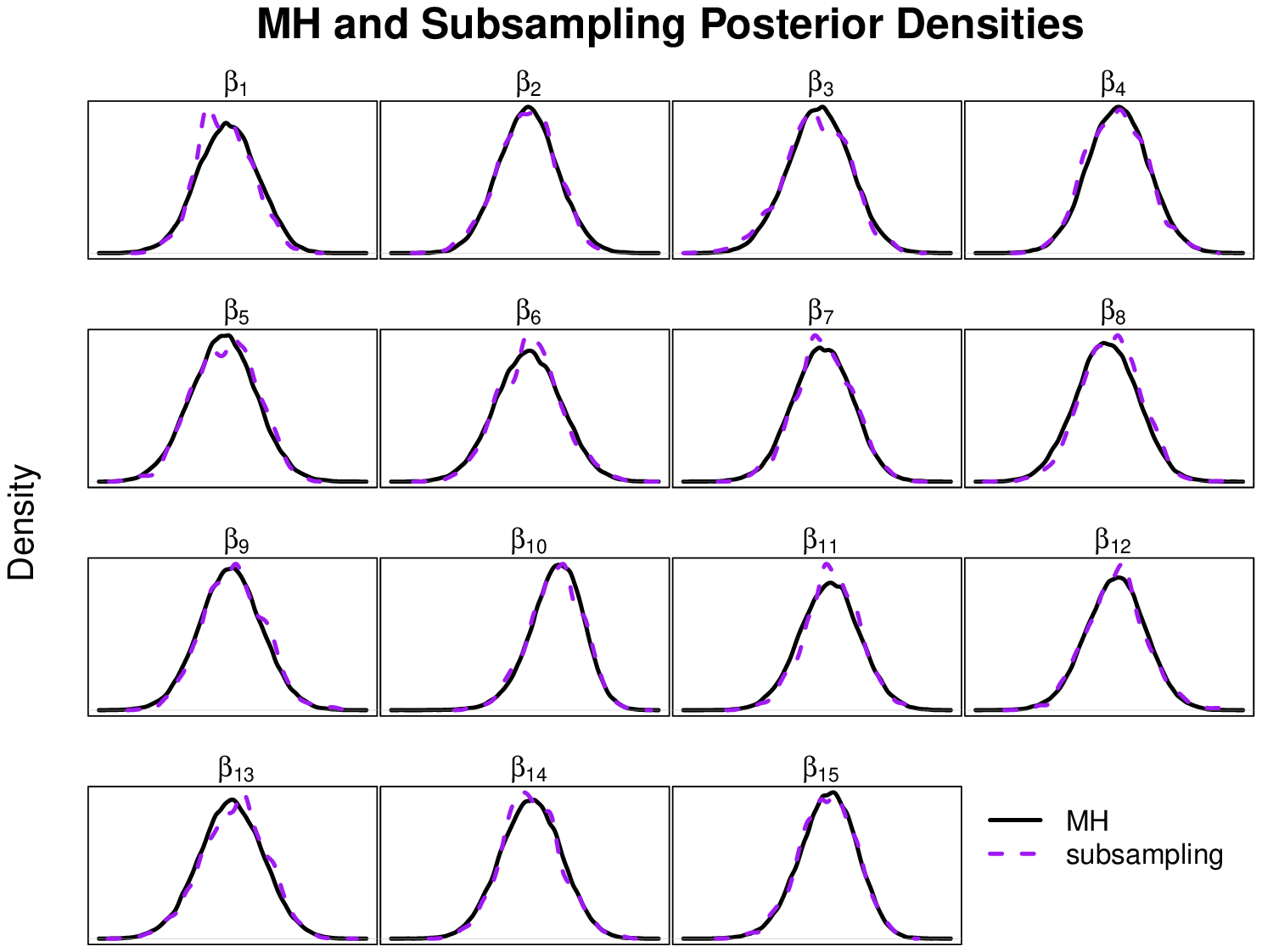}
\caption{Posterior densities from the Portuguese bank data, subsampling vs. MH. The MH and subsampling methods are represented by the solid and dashed lines, respectively.}
\label{fig:banksubsampling}
\end{figure}

\begin{figure}
\centering
%\captionsetup{format=hang}
\includegraphics[scale=.5,angle=270]{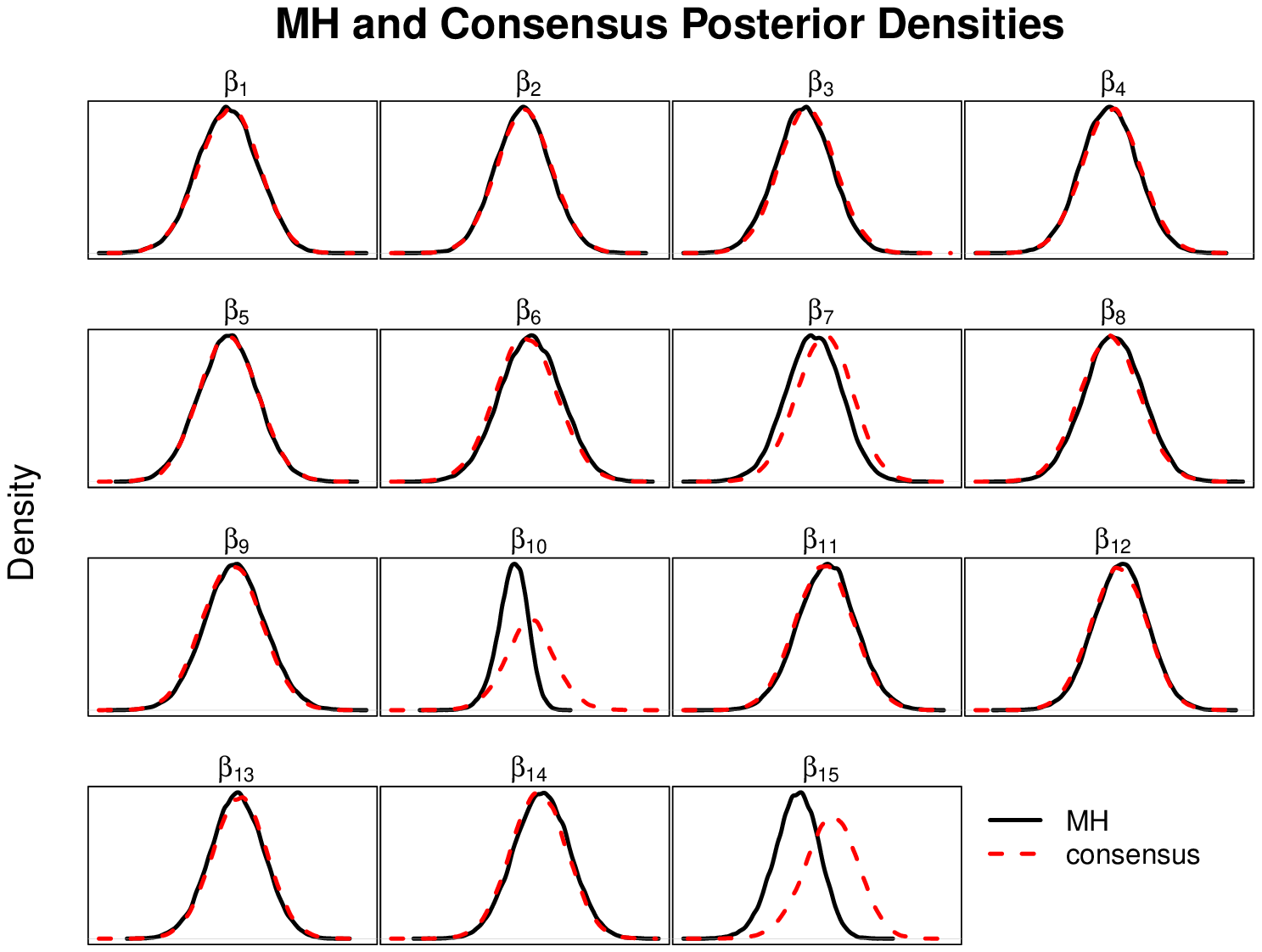}
\caption{Posterior densities from the Portuguese bank data, consensus Monte Carlo vs. MH. The MH and consensus Monte Carlo methods are represented by the solid and dashed lines, respectively.}
\label{fig:bankconsensus}
\end{figure}

Figures~\ref{fig:banktwostage}-\ref{fig:bankconsensus}
compare the posterior densities of the two-stage, subsampling and consensus methods to the standard Metropolis sampler results.
Figure~\ref{fig:banktwostage}
shows that the two-stage method matches the results obtained by the unmodified MH algorithm which is expected based on the theoretical results above.  Figure~\ref{fig:banksubsampling} 
indicates that the subsampling method was effective in capturing the true posterior distribution since the subsampling method can be arbitrarily close to the true posterior based on the subsample size~\citep{quiroz}. Figure~\ref{fig:bankconsensus}
shows that the consensus method matches the true posterior very well, with the exception of $\beta_{10}$ and $\beta_{15}$ which have a larger spreads and are slightly biased.  Interestingly, $\beta_{10}$ and $\beta_{15}$ are the coefficients for education (`illiterate'), and marital status (`unknown') which have only 18 and 80 cases in the dataset respectively.  Paradoxically, as ~\citet{scott} points out, we are suffering from a case of small sample bias in a large dataset, which is a potential issue in consensus Monte Carlo applications.
Since the Portuguese bank dataset is relatively small (approximately 40,000 observations), we refer the reader to the next section to better understand how these methods might scale to larger datasets, as well as a more detailed comparison of the speed and efficiency of the methods.

\subsubsection{Freddie Mac Data, Logistic Regression}
The loan data from Freddie Mac was obtained in September 2015 from Freddie Mac's website.  The data consists of approximately 2.3 million loans which Freddie Mac acquired during 2009-2010 and contains monthly performance data on each loan.  The binary dependent variable of interest is whether or not a loan was foreclosed by the end of September 2014.

In order to understand and quantify the effects of various covariates on foreclosure, a logistic model was used.  Covariates of interest include the date of the first mortgage payment, FICO score, debt to income ratio, original principal balance of the loan, and first-time home-buyer status (yes, no, unknown).  Each variable was transformed, centered, and scaled as appropriate.  A vague prior was placed on $\bbeta$ yielding the following logistic model:
\begin{eqnarray} \label{eq:fredmod1}
y_i \ |\ \bbeta, \bx_i &\sim& \text{Bernoulli}\{\pi(\bx_i)\} \nonumber \\
\pi(\bx_i) &=& \{1 + \text{exp}(-\bx_i\boldsymbol{\beta})\}^{-1}\\
\boldsymbol{\beta} &\sim& \text{Multivariate-Normal}(\mathbf{0},\Sigma_0 = I * 10^2) \nonumber
\end{eqnarray}
where $\Bbeta$ is the vector of coefficients, $\Sigma_0$ is the covariance matrix for $\Bbeta$, $I$ is the identity matrix of appropriate dimension and $\bx_i$ is the $i$th row of the design matrix, $i=1,\ldots, n$.

The coefficients of the model were estimated using the usual MH, consensus MH, and two-stage MH algorithm, all of which were coded in Fortran. In order to provide a fair comparison with the consensus Monte Carlo algorithm, both the MH and two-stage MH algorithm were parallelized with $p=14$ partitions as described in Section 2.1 using Open MPI for Fortran.  In the two-stage MH algorithm, the $s$ observations used to approximate the log-likelihood in the first stage were selected by randomly selecting $s/p$ observations from each partition prior to the start of the MH algorithm.

The subsampling MH method was not employed on the Freddie Mac dataset since it was not likely to computationally competitive in this setting.  Since the data are extremely sparse, the likelihood can be easily calculated for the cases when $y_i =1$, so subsampling would only need to be employed when $y_i = 0$.  For full implementation, three separate spline surfaces would be required to be fit for each category of first-time home-buyer status (no, yes, unknown).  Even if a relatively small sample was used for each spline surface  approximation (e.g.\ several thousand observations), the corresponding matrices to calculate the spline fits would be in total far larger than the design matrix itself, and that computation is only the first step.  Furthermore, the subsampling method is not likely to see the gains of parallelization that the MH and two-stage MH algorithms receive since more data will need to pass between processes and/or each process will have to calculate the same information in parallel (which defeats the purpose of parallelization).

\begin{figure}
\centering
%\captionsetup{format=hang}
\includegraphics[scale=.4,angle=270]{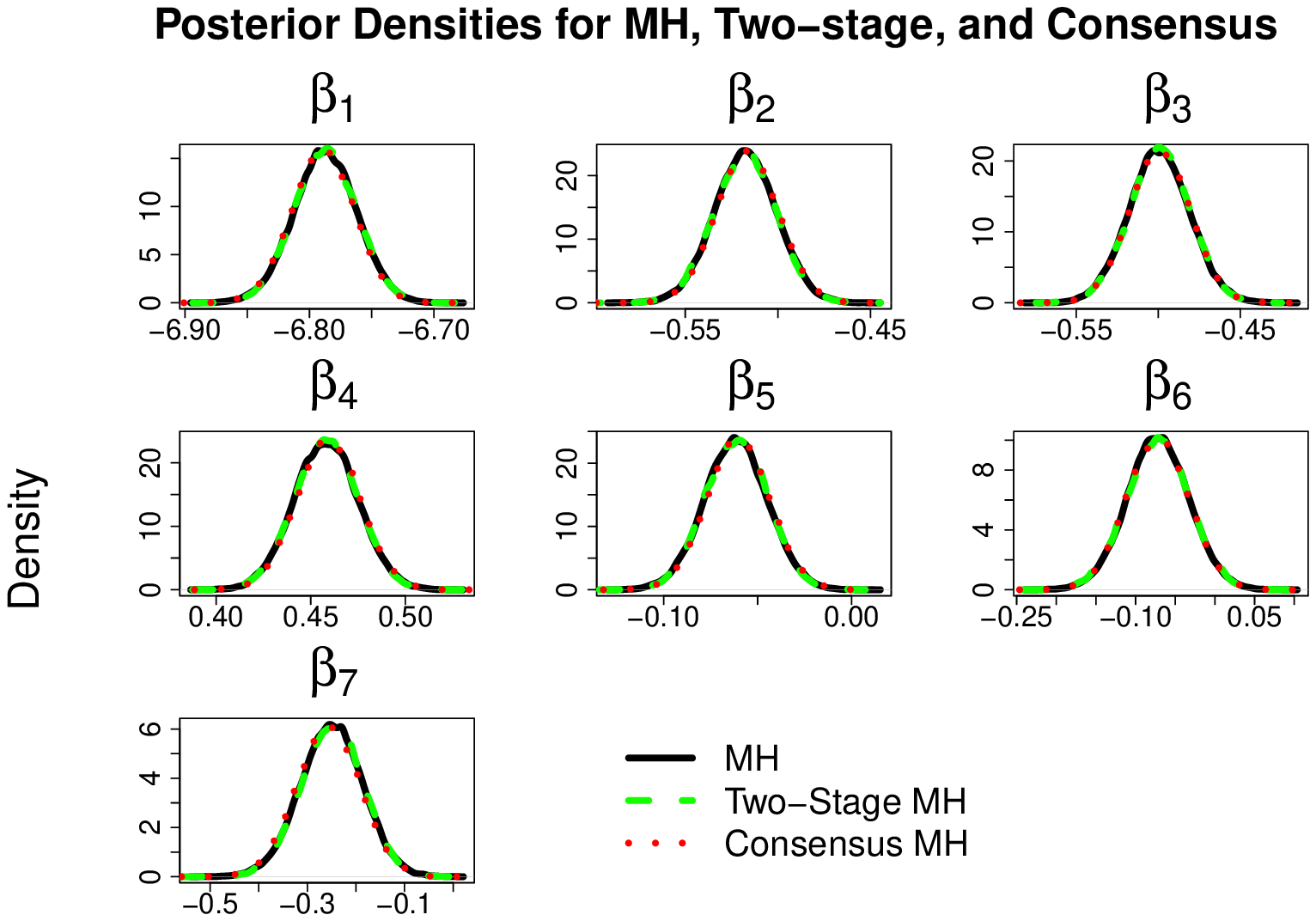}
\caption{Posterior densities from Freddie Mac data. MH, two-stage MH, and consensus MH.}
\label{fig:simplemod_densities}
\end{figure}

\begin{figure}
\centering
%\captionsetup{format=hang}
\includegraphics[scale=.5,angle=270]{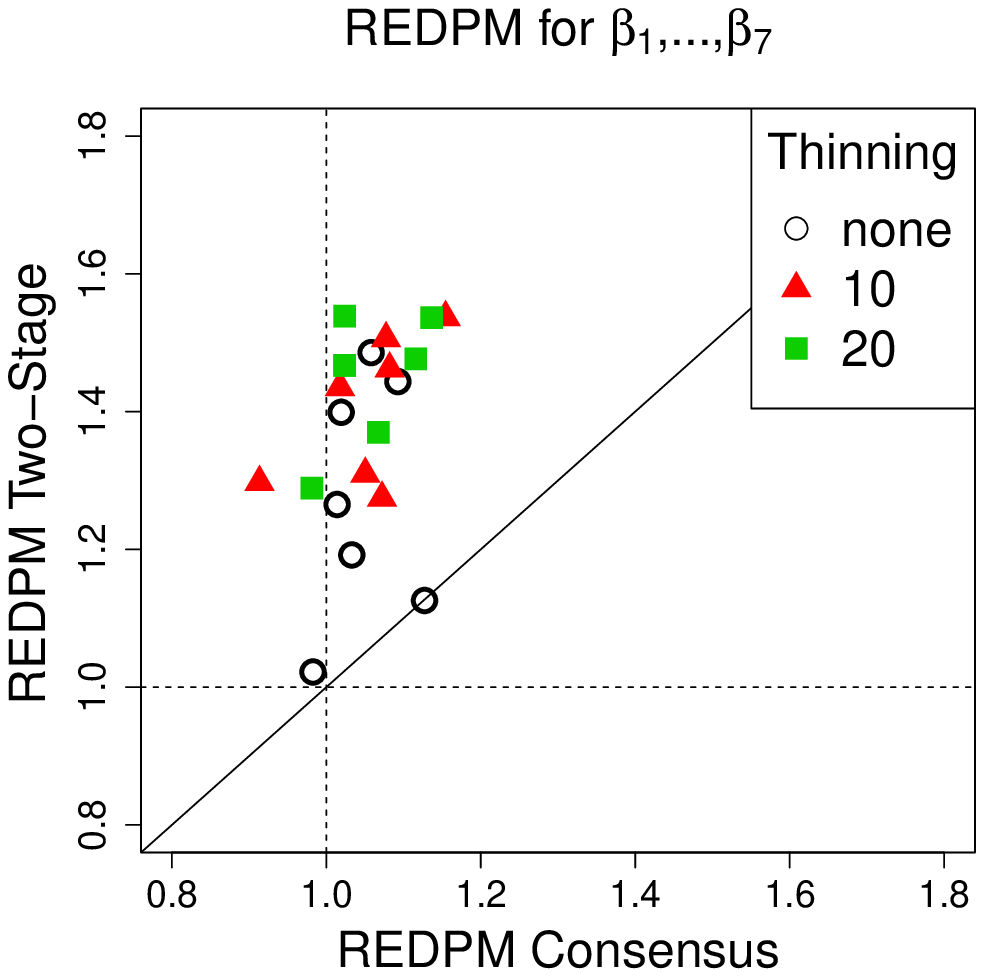}
\caption{REDPM with respect to the MH algorithm for two-stage and consensus MH.  REDPM is plotted for the original MCMC chain and the chain thinned every 10 and 20 values.}
\label{fig:redpm_simplemod}
\end{figure}

The usual MH, consensus, and two-stage MH algorithms were run for 100,000 iterations with a burn-in period of 5,000.  Parameters were updated sequentially and proposal variances were chosen such that the acceptance rate for each parameter was near 50\%. Figure~\ref{fig:simplemod_densities}
plots the densities of the posterior distribution of $\beta_1,\ldots,\beta_7$.  The densities are essentially indistinguishable between the MH, two-stage, and consensus methods.  The execution times were 118, 115, and 81 minutes for the parallelized MH, consensus, and two-stage methods, respectively.  In this particular application, the autocorrelation in the two-stage method was slightly more persistent than the regular MH algorithm.  Consequently, it is of interest to evaluate the efficiency of the three MCMC chains, accounting for autocorrelation.  This can be done by measuring the effective draws per minute (EDPM), which is a measure of the equivalent number of independent posterior draws per minute the MCMC chain represents.  The EDPM diagnostic incorporates both the execution time and autocorrelation of the chain to measure its efficiency:
\[
\text{EDPM} = t^{-1}\left(\frac{n}{1 + 2\sum_{k=1}^\infty \rho_k}\right)
\] 
where $n$ is the number of MCMC iterations, $t=$ execution time of the MCMC chain in minutes, and $\rho_k$ is the autocorrelation at the $k$th lag of the chain.  EDPM can be calculated by estimating $\rho_k$ with $\hat{\rho}_k$, the sample autocorrelation of the MCMC chain.  In order to compare the efficiency of two MCMC chains, we can compute the relative effective draws per minute (REDPM) as 
\[
\text{REDPM} = \frac{\text{EDPM}_{\text{Algorithm 1}}}{\text{EDPM}_{\text{Algorithm 2}}}
\]
Figure~\ref{fig:redpm_simplemod}
plots the REDPM of the two-stage and consensus methods relative to the MH method for each coefficient, $\beta_1,\ldots,\beta_7$.  Also plotted are the REDPM values for the MCMC chain thinned by keeping every 10th and 20th values of the chain.  We note that the two-stage method had REDPM values which were always above 1, and with the exception of one parameter, was always above the REDPM values of the consensus method.  The median REDPMs for the two-stage method were 1.27, 1.44, and 1.47 as contrasted with 1.03, 1.07, and 1.02 for the consensus method (no thinning, keeping every 10th and 20th observations respectively).  Thus in this application, the two-stage method appears to perform best in terms of speed, efficiency, and accuracy.
\subsubsection{Freddie Mac, Hierarchical Logistic Regression}
Bayesian statistics provide a simple way to fit hierarchical models, and with the help of MCMC, estimation of the parameters is generally straightforward.  In addition to the covariates in the previous logistic regression model, the Freddie Mac dataset specifies which bank originally serviced the loan.  It is of particular interest to understand how delinquency rates vary between banks during this time period.  In order to accomplish this, we specify the following hierarchical model:
\begin{eqnarray*}
y_{ij}\ |\  \theta_j, \bbeta, \bx &\sim& \text{Bernoulli}\{\pi(\bx_{ij})\}\\
\pi(\bx_{ij}) &=& [1 + \text{exp}\{-(\theta_j + \bx_{ij}\boldsymbol{\beta})\}]^{-1}\\
\boldsymbol{\beta} &\sim& \text{Multivariate-Normal}(\mathbf{0},\Sigma_0 = I * 10^2) \\
\theta_j &\stackrel{iid}{\sim}& \text{Normal}(0,\tau^2) \\
p(\tau) &\propto& \tau^{-2}
\end{eqnarray*}
where $\Bbeta$ is the vector of coefficients (the same covariates as in (\ref{eq:fredmod1}) with an intercept), $\Sigma_0$ is the covariance matrix for the vague prior on $\Bbeta$, $I$ is the identity matrix of appropriate dimension, $\bx_{ij}$ is the row of the design matrix corresponding to observation $y_{ij}$ and $\theta_j$ represents a random intercept term for the $j$th bank who serviced the loan, $j=1,\ldots,k=16$, $i=1,\ldots,n_j$.  Lastly, Jeffrey's prior was placed on $\tau$.

In this model, interest lies primarily in the posterior distribution of $\tau$, which provides us with an understanding of the variability of loan foreclosure rates between banks after controlling for the other covariates.  For datasets which have relatively small $n$, the MH algorithm is straightforward to implement on this simple hierarchical model.  The two-stage method is also easily extended to this hierarchical model, however, the consensus and the subsampling methods are not as easily implemented.

The two-stage and the usual MH algorithms were successfully implemented.  As before, the data was partitioned into 14 partitions and the likelihoods were computed in parallel on each iteration.  Both chains were run for 100,000 iterations with a burn-in period of 5,000.  Parameters were updated sequentially and proposal distributions were chosen such that the acceptance rates for each parameter was near 50\%.  The two-stage method was implemented twice with sample sizes of 224,000 and 22,400 observations which were sampled prior to running the algorithm (1000 and 100 data values from each bank on each partition, roughly 10\% and 5\% of data).  The total run times for the parallelized MH and the two-stage MH (10\% and 5\% subsample) were 1106, 849, and 639 minutes, respectively.  We note that the variance of the proposal distribution for the two-stage MH with 5\% subsampling was reduced (compared to the MH and two-stage MH with a 10\% sample) in order to obtain the desired acceptance rate of the MCMC chain.

\begin{figure}
\centering
%\captionsetup{format=hang}
\includegraphics[scale=.5,angle=270]{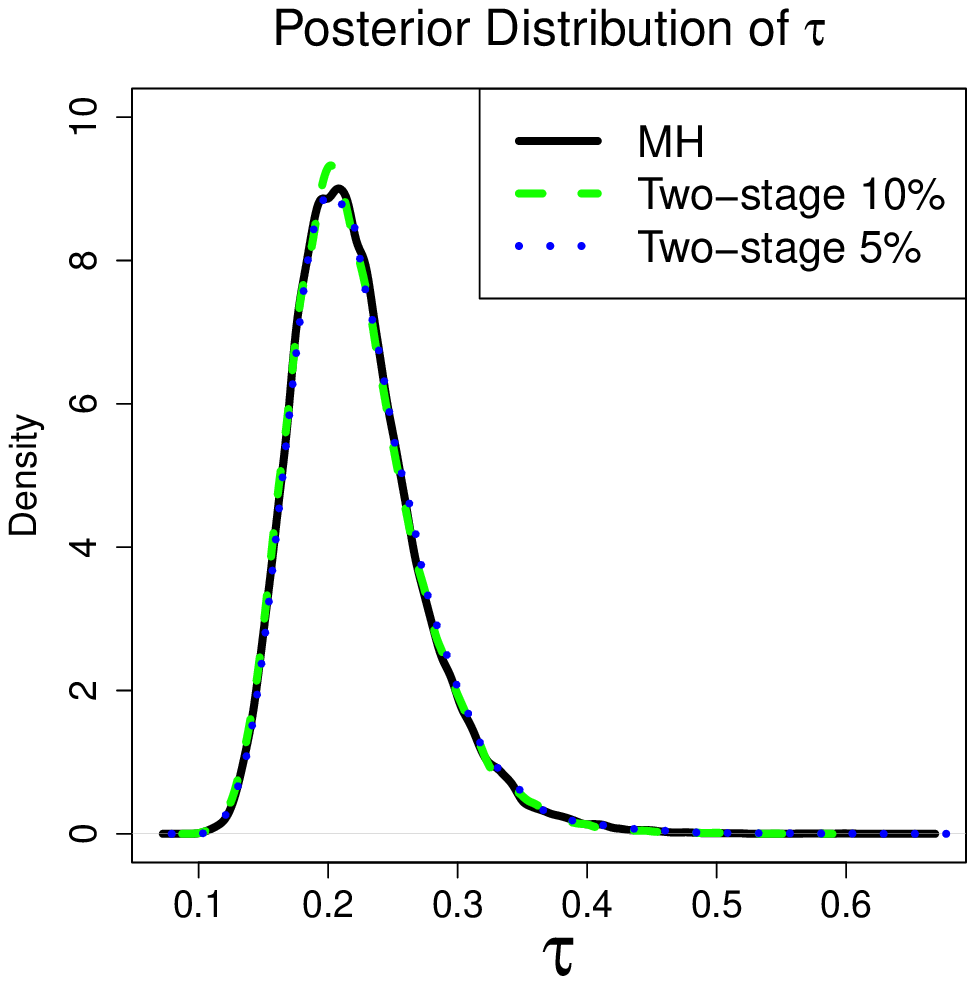}
\caption{Posterior Distribution of $\tau$ for the MH and two-stage method subsampling 5\% and 10\% of the data.}
\label{fig:tau}
\end{figure}

As in the other two applications, the posterior densities for the MH and two-stage MH for the 24 parameters were within Monte Carlo error (since both the MH and two-stage MH algorithm attain the correct stationary distribution).  For brevity, we plot only the posterior distribution of $\tau$ since it is the primary parameter of interest (Figure~\ref{fig:tau}).

\begin{figure}
\centering
%\captionsetup{format=hang}
\includegraphics[scale=.5,angle=270]{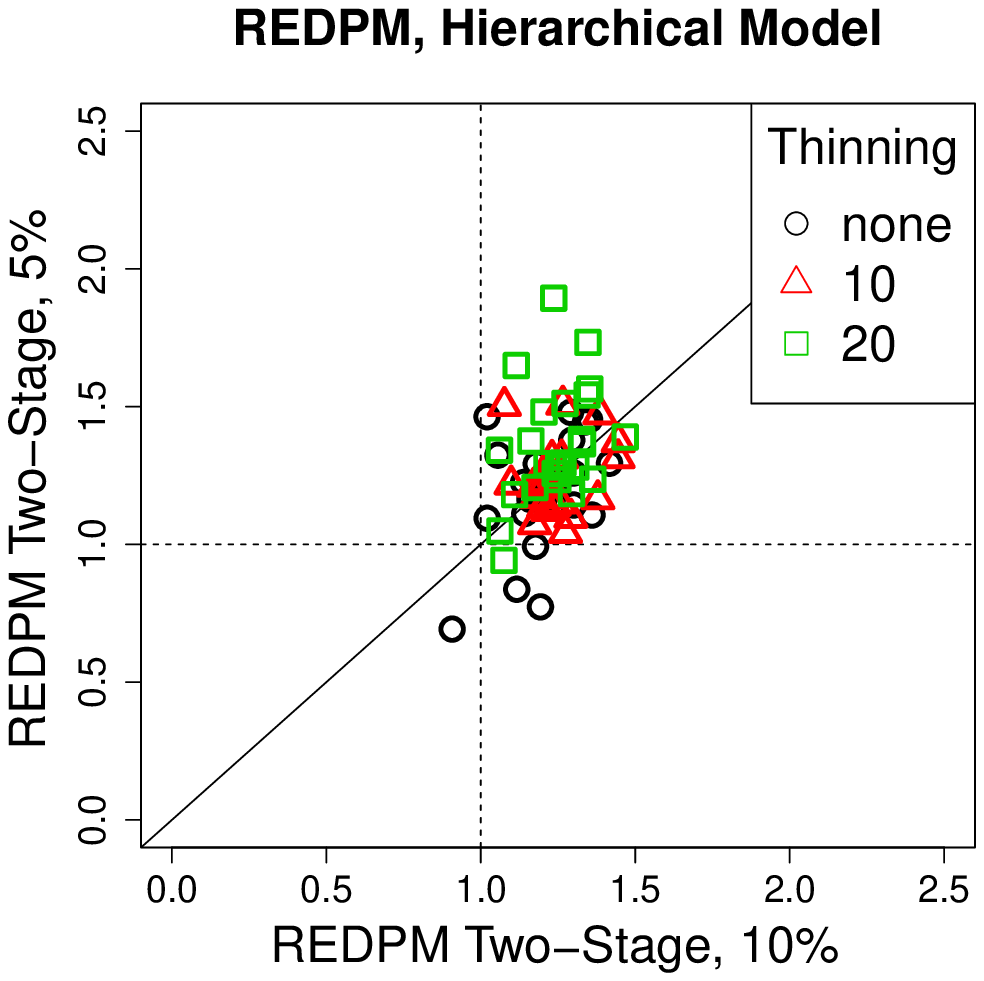}
\caption{REDPM for the two-stage method with respect to the MH algorithm for subsample sizes of 5\% and 10\%.}
\label{fig:redpm}
\end{figure}

Figure~\ref{fig:redpm} 
plots the REDPM values of the 24 parameters for the hierarchical model.  We note that the two-stage algorithm using 10\% of the data consistently had REDPM values greater than 1, whereas using 5\% of the data yielded more variability in the REDPM values, including 4 values lower than 1 on the un-thinned MCMC chain.  In the 5\% sampling case, all the values of REDPM $< 1$ were elements of $\Bbeta$, not $\Btheta$.  This may be an artifact of the sampling design since sampling was stratified by bank, and therefore some of the covariates may not have had adequate coverage in the smaller sample size.  Thinning improves REDPM most dramatically for low REDPM values in the 5\% sample, but otherwise doesn't seem to cause any major shifts in REDPM.  Overall, the two-stage method showed increases in efficiency for the majority of the parameters.

The consensus method can be applied to this model as long as all the loans originating from a particular bank are in the same partition.   The method proposed by ~\citet{scott} requires running independent MCMC chains in parallel and then combining the draws of $\tau$ in the usual manner, and then discarding the values of $\Btheta$ and $\Bbeta$.  Once the draws of $\tau$ are combined, these values are sent to each partition which independently draw new values of $\Bbeta$ and $\Btheta$ from $p(\Bbeta | \tau, \Bx)$ and $p(\Btheta | \tau, \Bx)$.  In our case, however, these distributions are not in standard form and are not easily sampled from.  In implementation, the first consensus MCMC chain to obtain draws of $\tau$ was faster than the traditional MH algorithm with a parallelized likelihood but performed more slowly than the two-stage method (1106, 910, 849, 639 minutes for MH, consensus, and two-stage MH (10\% and 5\% subsampling) respectively).   Since the speed of the first run of the consensus method was slower than the two-stage method and the two-stage MH was more efficient than the consensus method in the previous model, drawing values from $p(\Bbeta | \tau, \Bx)$ and $p(\Btheta | \tau, \Bx)$ was not implemented.

The subsampling method can also in theory be applied to this model.  However, this requires fitting 48 spline surfaces prior to running the MCMC (16 banks, 3 levels of first-time home-buyer status).  These spline surfaces were fit using the methodology provided by~\citet*{ma2015spline} using a subsample of $s=16,000$ observations (1,000 observations per group).  However, on each iteration, approximating the log-likelihood surface for the entire dataset requires 48 matrix multiplications of dimension $z_i \times s, \sum_{i=1}^{48} z_i = n - s - n_1 = 2,297,813 - 3,711 - 16,000 = 2,278,102$ where $n_1 = \sum I(y_i = 1)$.  These matrices were too large to fit into RAM, thus we were unable to implement the subsampling MH.  Even if the data did fit into RAM, the computational cost of estimating the likelihood contribution with splines would likely be greater than evaluating the likelihood directly.  Furthermore, implementing the subsampling method in parallel is not likely to produce significant gains in computation time since it will require either calculating the same quantities on each process (which defeats the purpose parallelizing) or passing vectors of information (rather than scalars) between processes.

\subsection{Bayesian MARS}
The two-stage method also has applications in more complicated classification settings, including Bayesian multivariate adaptive regression splines (BMARS) \citep*{friedman1991multivariate,holmes2003classification}.  BMARS is a non-linear classification method which is extremely flexible for classification problems where the relationship between the response and covariates is complex, unknown, or otherwise difficult for the analyst to specify. It uses the data to adaptively choose splines and knots to flexibly model classification problems.   Since the splines and knot locations are not known a priori, BMARS requires the use of reversible jump MCMC \citep*{green1995reversible} to explore a parameter space with varying dimension.

Even in this more complicated setting, implementing the two-stage MH requires only a few extra lines of code, but can still produce a faster MCMC chain.  To quantify the effectiveness of the two-stage method using BMARS, one million observations were simulated from the following model:
\begin{eqnarray*}
&y \sim \text{Bernoulli}\{ [1 + \exp(-\pi(\mu))]^{-1}\} \\
&\mu = x_1 + x_2 - x_3 - x_4 + x_1x_2  - .5x_1x_3 - x_2 x_3 + .2 x_1  x_2  x_3; \\
& x_1 \sim U(0,1),\ x_2 \sim N(0,1),\ x_3 \sim U(0,2),\ x_4 \sim N(0,2^2)
\end{eqnarray*}
Where $U(a,b)$ denotes a uniform distribution on the interval $(a,b)$. The BMARS method was implemented using the prior distributions outlined by \citet*{holmes2003classification}, to which we refer the reader to their paper for details. The two-stage method was implemented by choosing a random subsample prior to MCMC which was used in each iteration to approximate the likelihood.  The log-likelihood approximation in the first stage was calculated as $\hat{l} =  (n / a) l_{sub}$ where $n$ and $a$ are the number of observations in the whole dataset and subsample, respectively, and $l_{sub}$ is the log-likelihood contribution of the subsample.

The BMARS algorithm was run a total of 10 times on this simulated dataset.  The usual BMARS algorithm was run 5 times and the average is summarized in the first row of Table~\ref{tab:BMARS}.  The two-stage method was implemented on the remaining five runs with various subsampling percentages.  Once the priors are in place, there are only two parameters which need to be specified in the BMARS method: the maximum number of interactions allowed for the basis functions (which was chosen to be 3), and a tuning parameter (the proposal standard deviation of the spline coefficients).  

From Table~\ref{tab:BMARS}, it is clear that the two-stage method is faster than the usual BMARS MCMC, with all two-stage runs producing a 30\%-40\% reduction in time.  As with the previous examples, as the subsampling percentage decreased, the acceptance rate of the two-stage MCMC chain decreased and the speed increased.  When sampling only one percent of the data, the acceptance rate was very low, so it was re-run with a smaller proposal standard deviation.  This led to an increase in the acceptance rate with a slight reduction in speed.

\begin{figure}
\centering
%\captionsetup{format=hang}
\includegraphics[scale=.8]{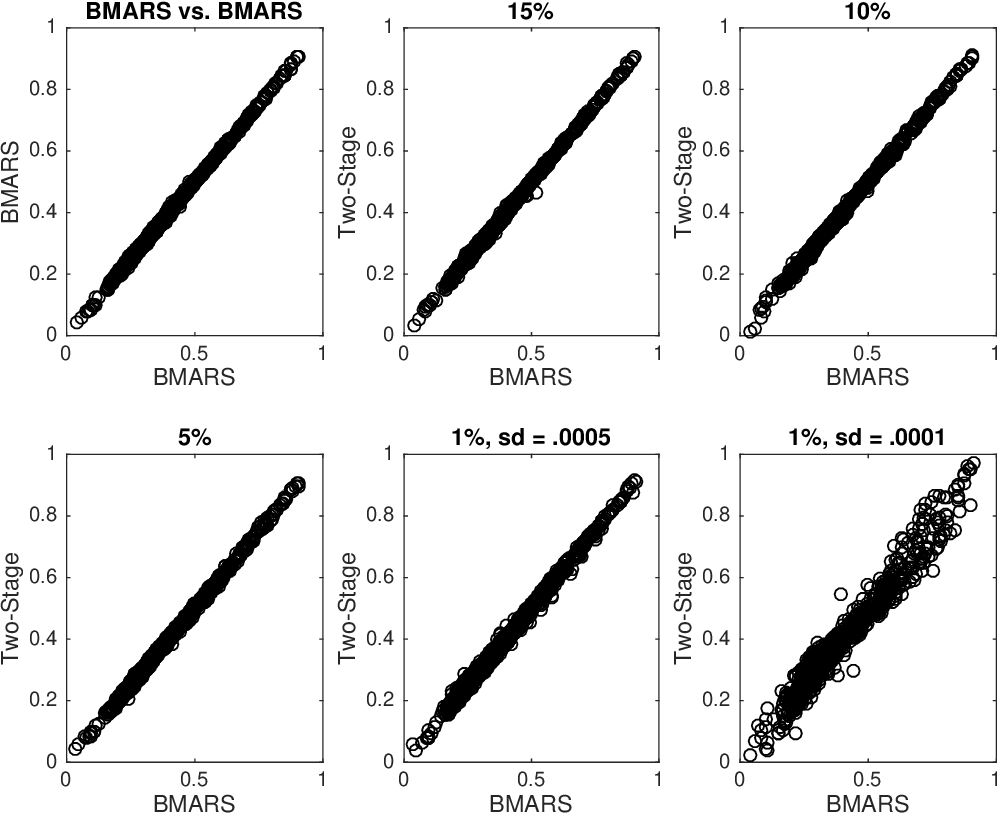}
\caption{Comparison of the test-set predictions between the BMARS MCMC and the two-stage BMARS MCMC for various subsampling percentages.  The top-left panel compares the predicted probabilities between two BMARS runs for a visual of Monte Carlo error. }
\label{fig:BMARS}
\end{figure}

Due to the varying dimension of the parameter space during MCMC, comparing efficiency of the MCMC chain is not straightforward.  Consequently, to determine the effectiveness of the two-stage method, one thousand observations were used as a test set, and the predictions based on both MCMC chains were compared and are shown in Figure~\ref{fig:BMARS}.  The top-left panel of Figure~\ref{fig:BMARS} compares the predicted probabilities of two BMARS runs (neither implementing the two-stage method) to provide a visual of Monte Carlo error.  The two-stage method with 15\%, 10\%, and 5\% subsampling produced predictions which appear to be within Monte Carlo error of the usual BMARS algorithm.  The two-stage 1\% subsampling (sd = .0005) showed slightly more variability and 1\% subsampling (sd = .0001) shows a fair amount of variability in the predictions.  Interestingly, although the predictions between the BMARS and two-stage methods become more variable as the subsampling percentage decreased, the root-mean-square error (RMSE) of the predictions were essentially the same (RMSE for the 5 BMARS runs: .0806, .0805, .0806, .0806, .0807; RMSE for the 5 two-stage runs: .0807, .0806, .0808, .0804, .0806 for 15\%, 10\%, 5\%, 1\% (sd = .0005), 1\% (sd = .0001), respectively).   This gives further evidence that the two-stage method can still be highly effective even when using a very small percentage of the data as a subsample.

\input{Table1}

\section{Discussion}
Perhaps the most pressing question regarding two-stage MH is how to select the subsample size.  From experience the authors note that for a fixed  proposal distribution variance, decreasing the subsampling percentage will at some point decrease acceptance rates of the MCMC chain.  This is due to the fact that the estimate of the likelihood is either overestimating or underestimating the likelihood ratio which causes proposed parameter values to be discarded by either the first stage (if the estimate of the likelihood ratio is too small) or the second stage (if the estimate of the likelihood ratio is too large).  If too small of a subsample is used, the variance of the proposal distribution will need to be reduced to obtain the desired acceptance rate of the MCMC chain.  A smaller subsample will increase the speed of the chain, but will likely increase the autocorrelation of the chain since the variance of the proposal distribution will need to be reduced.  Even so, the hierarchical model for the Freddie Mac data still performed well with sampling only 5\%-10\% of the data, and the BMARS application performed well even with 1\%-15\% subsampling.  This indicates that the speed and efficiency of the two-stage method may be somewhat robust to the subsample size used to approximate the log-likelihood. 

One of the main advantages of the two-stage method is its simplicity and ease of implementation.  It requires taking only one subsample prior to the MCMC algorithm and then adding a few lines of code to implement the first screening stage.  Furthermore, it can be applied to any model in which a computationally cheap estimate of the likelihood can be obtained. Even using naive likelihood approximations, the two-stage method has performed well.  If more precise likelihood estimates can be acquired for a particular model, the two-stage method may be even more effective at screening out bad proposals (although the speed will still depend on a computationally cheap likelihood estimate).

The consensus method is also generally straightforward in simple models, but even in hierarchical models it places restrictions on how the data can be partitioned and may require sampling from distributions which cannot be sampled from directly (which adds another potentially computationally demanding layer).  The subsampling method requires the most effort to implement since it requires fitting spline surfaces to the data.  Furthermore, these spline surfaces may require very large matrix multiplications to provide the approximation to the likelihood surface on each iteration of the MCMC.

The success of the two-stage method on the complex BMARS method indicates that it has potential in many other applications.  Other potential non-linear classification methods include relevance vector machine \citep*{tipping2001sparse} and support vector machine models \citep*{mallick2005bayesian}. This can also be extended in a multivariate responses framework \citep*{holmes2003generalized}.  Perhaps most importantly, the two-stage method is not limited to classification problems.  It can be applied to any model where a computationally cheap and accurate approximation of the likelihood can be constructed.

\section{Conclusion}
The results from this paper indicate there are a number of tall data Bayesian methods which are effective in obtaining/approximating the posterior distribution more quickly than traditional methods.  Two-stage MH is simple to implement, fast, and overall more efficient than consensus, subsampling, or unmodified MH algorithms in our applications.  Combining two-stage MH with the consensus method shows promise for even larger datasets in which the data cannot fit in RAM. Future extensions to this work include applying the method to handle more complicated likelihoods, and finding better likelihood approximations which are still computationally cheap to evaluate.

\section{Acknowledgments}
The authors would like to thank the reviewers for their comments which greatly improved the presentation of the material in this article.

%\bibliographystyle{classificationstyle3}
%\bibliography{paperref}

\end{spacing}
\end{document}

%% file: Table1.tex
\begin{table}
\begin{center}
\begin{tabular}{ccccc}
Subsample Percentage & SD & Acceptance Rate & Time (sec) & Time Ratio\\
- & .0005 & .31 & 41,594 & - \\
15 & .0005 & .18 & 28,134 & .68 \\
10 & .0005 & .15 & 27,422  & .66 \\
5 & .0005 & .12 & 25,384 & .61 \\
1 & .0005 & .06 & 25,144 & .60\\
1 & .0001 & .21 & 28,261 & .68 \\
\end{tabular}
\end{center}
\caption{The results of the BMARS MCMC with the two-stage BMARS MCMC.  The first row corresponds to the average of 5 runs of the usual BMARS algorithm to which the two-stage runs are compared.  The last column is the ratio of the two-stage time to the regular MCMC time in the first row.  Note that as the subsampling percentage decreased, the speed of the two-stage algorithm increased while the acceptance rate decreased for a fixed proposal standard deviation (SD).}
\label{tab:BMARS}
\end{table} 